\documentclass{tcibook}
\usepackage{fancyhea}
\usepackage{work}
\usepackage{bm}       %    enables bold math symbols  e.g.  \bm{\gamma}
\usepackage{graphicx}
\usepackage{hyperref}      % hypertext links %%ARXIV
\usepackage{amsmath}
\usepackage{chngcntr}
\counterwithout{figure}{chapter}

\usepackage[usenames,dvipsnames]{color} %used for font color

\newcommand{\nc}{\newcommand}  

\def\beq{\begin{equation}}
\def\eeq#1{\label{#1}\end{equation}}
\def\eeqn{\end{equation}}

\newenvironment{Eqnarray}%
   {\arraycolsep 0.14em\begin{eqnarray}}{\end{eqnarray}}
\def\beqa{\begin{Eqnarray}}
\def\eeqa#1{\label{#1}\end{Eqnarray}}
\def\eeqan{\end{Eqnarray}}

\nc{\ra}{\rightarrow}  
\nc{\slsh}{\slash\hspace*{-0.22cm}}
\def\Re{{\cal R \mskip-4mu \lower.1ex \hbox{\it e}\,}}
\def\Im{{\cal I \mskip-5mu \lower.1ex \hbox{\it m}\,}}

\nc{\vev}[1]{ \left\langle {#1} \right\rangle }
\nc{\bra}[1]{ \langle {#1} | }
\nc{\ket}[1]{ | {#1} \rangle }
\nc{\fb}{\,{\rm fb}^{-1}}
\nc{\ev}{{\rm eV}}
\nc{\kev}{{\rm keV}}
\nc{\Mev}{{\rm MeV}}
\nc{\gev}{{\rm GeV}}
\nc{\tev}{{\rm TeV}}
\nc{\mev}{{\rm MeV}}

\def\del{\partial}
\def\Dslash{\not{\hbox{\kern-4pt $D$}}}
\def\dslash{\not{\hbox{\kern-2pt $\del$}}}
\def\pslash{\not{\hbox{\kern-2pt $p$}}}
\def\ETmiss{ \not{\hbox{\kern-4pt $E$}}_T }

\def\One{\ensuremath{\bf 1}\xspace}

\def\msb{{\bar{\ssstyle M \kern -1pt S}}}

\def\eps{\epsilon}

\newcommand{\cmbexp}{{CMB-S4}}

\begin{document}

\def\bibname{References}

\bibliographystyle{utphys}  %%%% MODIFIED FOR CF %%%%

\raggedbottom

\pagenumbering{roman}

\parindent=0pt
\parskip=8pt
\setlength{\evensidemargin}{0pt}
\setlength{\oddsidemargin}{0pt}
\setlength{\marginparsep}{0.0in}
\setlength{\marginparwidth}{0.0in}
\marginparpush=0pt

% The content begins here

\pagenumbering{arabic}

\renewcommand{\chapname}{chap:intro_}
\renewcommand{\chapterdir}{.}
\renewcommand{\arraystretch}{1.25}
\addtolength{\arraycolsep}{-3pt}

%%%% Author comment macros here

\def\as#1{[{\bf AS:} {\it #1}] }

%%%%%%%%%%%%%%%%%%%%%%%%%%%%%%%%%%%%%%%%%%%%%%%%%%%
%%%%%%%%%%%%%%%%%%%%%%%%%%%%%%%%%%%%%%%%%%%%%%%%%%%
%%%     All of your files should be in a subdirectory.  Here the
%%%     subdirectory is called CF0. The title of your
%%%     report should be wgreportCF0.tex in that subdirectory.  Input
%%%     that file here
%%%%%%%%%%%%%%%%%%%%%%%%%%%%%%%%%%%%%%%%%%%%%%%%%%%%
%%%%%%%%%%%%%%%%%%%%%%%%%%%%%%%%%%%%%%%%%%%%%%%%%%%

%%%%%% Inflation from CMB + LSS Document  %%%%%%%%%%%%%%%%

\def\la{\langle}
\def\ra{\rangle}
\def\beq{\begin{equation}}
\def\eeq{\end{equation}}
\def\d{\partial}
\def\const{\mbox{const}}
\def\e{{\rm e}}
\def\al{\alpha}
\def\eps{\varepsilon}
\def\d{\partial}
\def\l{\left(}
\def\r{\right)}
\def\la{\langle }
\def\ra{\rangle }
\def\k{{\bf k}}
\def\gtrsim{\raise-.75ex\hbox{$\buildrel>\over\sim$}}
\def\muk{\mu \mathrm{K}}
\def\sqdeg{\mathrm{deg}^2}
\def\fnl{f_\mathrm{NL}}
\def\fsky{f_\mathrm{sky}}
\def\Mp{M_P}
\def\planck{\textit{Planck}}

\chapter*{Inflation Physics from the Cosmic Microwave Background and Large Scale Structure}
\renewcommand*\thesection{\arabic{section}}

%%%%%%%%%%%%%%%%%%%
\begin{center}\begin{boldmath}

% list of authors

%\hyphenpenalty 10000

\begin{center}

\begin{large} {\bf  Topical Conveners: J.E.~Carlstrom,  A.T.~Lee} \end{large}

K.N.~Abazajian,
K.~Arnold,
J.~Austermann, 
B.A.~Benson,
C.~Bischoff,
J.~Bock,
J.R.~Bond, 
J.~Borrill,
I.~Buder,
D.L.~Burke,
E.~Calabrese, 
J.E.~Carlstrom,
C.S.~Carvalho, 
C.L.~Chang,
H.C.~Chiang, 
S.~Church,
A.~Cooray,
T.M.~Crawford$^*$,
B.P.~Crill, 
K.S.~Dawson,
S.~Das,
M.J.~Devlin,
M.~Dobbs, 
S.~Dodelson,
O.~Dor\'e, 
J.~Dunkley, 
J.L.~Feng,
A.~Fraisse,
J.~Gallicchio, 
S.B.~Giddings, 
D.~Green,
N.W.~Halverson,
S.~Hanany,
D.~Hanson, 
S.R.~Hildebrandt,
A.~Hincks, 
R.~Hlozek, 
G.~Holder, 
W.L.~Holzapfel,
K.~Honscheid,
G.~Horowitz,
W.~Hu, 
J.~Hubmayr, 
K.~Irwin, 
M.~Jackson,
W.C.~Jones, 
R.~Kallosh,
M.~Kamionkowski,
B.~Keating,
R.~Keisler,
W.~Kinney,
L.~Knox,
E.~Komatsu, 
J.~Kovac,
C.-L.~Kuo,
A.~Kusaka,
C.~Lawrence,
A.T.~Lee,
E.~Leitch, 
A.~Linde,
E.~Linder,
P.~Lubin,
J.~Maldacena,
E.~Martinec,
J.~McMahon, 
A.~Miller,
V.~Mukhanov, 
L.~Newburgh, 
M.D.~Niemack,
H.~Nguyen,
H.T.~Nguyen,
L.~Page, 
C.~Pryke,
C.L.~Reichardt,
J.E.~Ruhl, 
N.~Sehgal, 
U.~Seljak,
L.~Senatore,
J.~Sievers,
E.~Silverstein,
A.~Slosar,
K.M.~Smith, 
D.~Spergel, 
S.T.~Staggs, 
A.~Stark,
R.~Stompor,
A.G.~Vieregg,
G.~Wang, 
S.~Watson,
E.J.~Wollack,
W.L.K.~Wu,
K.W.~Yoon,
O.~Zahn, 
and M.~Zaldarriaga

\end{center}

\textsuperscript{*} Corresponding author. Tel.: 773-702-6452. Email address: tcrawfor@kicp.uchicago.edu.\\[0.2cm]

%\hyphenpenalty 1000

%Conveners are also listed separately in authorlist.tex

\end{boldmath}\end{center}

\section*{Executive Summary}
\label{sec:intro}
Precision cosmological measurements push the boundaries
of our understanding of the fundamental physics that governs our universe.  
In the coming years, cosmologists will be in a position to make major breakthroughs in our understanding of the
physics of the very early universe 
and be able to probe particle physics and gravity at the highest energy scales yet accessed. 
A major leap forward in the sensitivity of 
cosmological experiments is within our technological reach, 
leveraging past and current experience to tackle some of the most interesting fundamental physics questions.

Cosmic inflation, the theory that the universe underwent a violent,
exponential expansion during the first moments of time, is the leading
theoretical paradigm for the earliest history of the universe and for
the origin of the structure in the universe.  Current measurements of
the cosmic microwave background (CMB) and observations of the large-scale 
distributions of dark matter and galaxies in the universe are in
stunning agreement with the concept of inflation.  The next
generations of experiments in observational cosmology are poised to
decide central questions about the mechanism behind inflation.  In
this short document, we highlight the importance of experimentally
determining the nature of inflation in the early universe and the
unique opportunity these experiments provide to explore the physics of
space, time, and matter at the highest energies possible: those found
at the birth of the universe.

Although the landscape of possible models for inflation is potentially
large---and sensitive to quantum gravity corrections to the low-energy
quantum field theory---the phenomenology is sufficiently well
understood to make concrete distinctions between fundamentally
different classes of models that we can test observationally.

One key and generic prediction is the existence of a background of
gravitational waves~\cite{Starobinsky:1979ty} from inflation that produces a distinct signature
in the polarization of the CMB, referred to as ``$B$-mode''
polarization.  The amplitude of primordial gravitational waves, or
tensor modes, which can be detected or constrained by observations of
the $B$-mode polarization in the CMB, is fundamentally interesting for
several basic reasons.  It is proportional to the energy scale of
inflation and tied to the range of the inflaton field.  In particular,
observations promise to reach the level of sensitivity that will
enable them to determine whether the field range is larger than the
Planck scale in the simplest versions of inflation \cite{Lyth:1996im}.
This provides a striking ultraviolet-sensitive probe of quantum field
theory and quantum gravity, and an observational test of string
theoretic large-field inflation.  Additionally, in one 
theoretically developed (though currently speculative) alternative to inflation, the
ekpyrotic scenario, the authors of 
\cite{2002PhRvD..65l6003S,2001PhRvD..64l3522K} find no
mechanism for generation of the tensor perturbations;
hence, if these calculations are correct, detection of $B$-modes would present a convincing refutation of
that model.  Last but not least, a detection of tensor modes would
constitute a stunning measurement of the quantum
mechanical fluctuations of the gravitational
field.  

This motivates a next-generation CMB experiment with the sensitivity
and systematics control to detect such a polarized signal at
$\ge5\,\sigma$ significance, thus ensuring either a detection of
inflationary gravitational waves or the ability to rule out large
classes of inflationary models.  A program to meet these goals by
developing a Stage IV CMB experiment, \cmbexp, with $O$(500,000)
detectors by 2020 is described in the companion cosmic frontier
planning document 
{\it Neutrino Physics from the Cosmic Microwave Background and Large Scale Structure} \cite{Abazajian:2013oma}.
Such an experiment would
also contribute to inflationary science by strongly constraining the
spectrum of primordial density fluctuations, allowing us to
distinguish different families of inflationary models.

Possibilities for self-interactions of the inflaton and for additional fields are tested by different limits of the correlation functions of the perturbations. 
Despite important recent progress, we require substantial improvements before observational constraints on these quantities limit the interactions to be small corrections to slow-roll, or to detect non-Gaussianity if it is present.  A concerted theoretical effort combined with observations of large scale structure promises to fill this gap.    
A detection of primordial non-Gaussianity of the so-called local shape would effectively rule out all models
of inflation that involve a single scalar field \cite{Maldacena:2002vr,Creminelli:2004yq,bartolo04}. 
The CMB bound on local-model 
non-Gaussianity is now limited by having only one sky to observe; further improvements
will come from measurements of the large scale structure of the universe. The next 
generation of large scale structure measurements will produce non-Gaussianity constraints that are 
an important cross-check of the CMB bound and will pave the way for more stringent bounds
from future large scale structure measurements.

\section{Inflation Science:  theoretical motivations}
%%%%%%%%%%%%%%%%%%%
\label{sec:theory}
Cosmic inflation, the idea that the universe underwent a period of exponential expansion
in the first $10^{-34}$~seconds of its existence, was 
proposed in the early 1980s to explain the apparent smoothness and flatness of the universe
and the absence of relics such as magnetic monopoles \cite{Guth}.
Quantum fluctuations generated during inflation evolve into the distributions of dark matter and galaxies we observe today~\cite{Mukhanov:1981xt}. Inflation drives the spatial curvature to nearly zero, and introduces density perturbations that are adiabatic with a nearly scale-invariant spectrum that depends on the details of the inflationary potential.  

The frontier of inflation research currently lies in measurement of the polarization of the CMB and in searching for non-Gaussianity in the distribution of dark matter and galaxies in the late universe. The CMB offers a unique window between the late-time universe dominated by dark matter and dark energy, and the early universe when the energy density was dominated by the potential that drove cosmic inflation. The amplitude of tensor $B$-mode polarization in the CMB is proportional to the energy of inflation and tied to the range of the inflaton field. The rich phenomenology of non-Gaussianity in the distribution of dark matter and galaxies in the late universe offers opportunities to directly study the dynamics of inflation.  

In the context of inflationary paradigm, we can be precise about the significance and interpretation of these measurements.  Here we briefly summarize some highlights.   

The predictions of most inflationary models can be characterized in terms of the statistical properties of perturbations to the metric away from a homogeneous background solution $a(t)\approx e^{Ht}$.
We can parameterize the metric as
\begin{equation}
ds^2=-N^2dt^2+h_{ij}(dx^i+N^idt)(dx^j+N^jdt)
\end{equation} 
where
\begin{equation}
h_{ij}=a(t)^2\left[e^{2\zeta}\delta_{ij}+\gamma_{ij}\right]
\end{equation}
and $N, N^i$ represent the lapse and shift, non-dynamical modes of the metric that enforce constraints.
Here $\zeta$ contains the scalar perturbation (in a gauge where the inflaton perturbation has been gauged away via time reparameterization), and $\gamma$ the tensor perturbation.   The CMB and LSS are sufficiently linear in the regimes of interest that these primordial metric perturbations can by inferred directly from observations.  The greater challenge is to make inferences about the physics of inflation from knowledge of $\zeta$ and $\gamma_{ij}$.

\subsection{Tensor Modes}

Determining the tensor to scalar ratio 
\beq\label{rthreshold}
r=\frac{\langle \gamma\gamma\rangle }{\langle\zeta\zeta\rangle}
\eeq
via a measurement of the primordial $B$-mode polarization \cite{Seljak:1996gy, Kamionkowski:1996zd, Farhang:2011ud, Baumann:2008aq} is important for three simple reasons.\footnote{Exceptions to these going beyond single field slow-roll inflation, for example postulating rapid variation of the slow-roll parameter $\dot H/H^2$ \cite{BenDayan:2009kv, Hotchkiss:2011gz}, or including gravitational waves from sources produced during inflation \cite{Senatore:2011sp}, bring in their own motivations on par with that derived from the relation (\ref{Lyth}) in the simplest cases.  Although these more exotic possibilities are interesting, with space constraints we will not lay out the various caveats.}

\smallskip

\noindent (1)  A detection would constitute a measurement---for the first time---of the quantum mechanical fluctuations of the metric:  in the absence of classical inhomogeneities ($\langle\gamma_{ij}\rangle=0$) inflation generates a nonzero variance 
\beq\label{tensortwo}
\langle\gamma_{s, k} \gamma_{s', k'}\rangle'= \frac{1}{2 k^3 } \ 2 \, \frac{H^2}{\Mp^2} \,  \delta_{ss'}\delta^{(3)}(\k + \k^{'})  \ ,
\eeq  
where $s,s'$ label graviton polarizations and $\langle\ldots \rangle'$ denotes dropping the momentum conserving delta function.    

\smallskip

\noindent (2)  The formula (\ref{tensortwo}) also exhibits a direct connection between the tensor signal and the scale of inflation ($H$ or equivalently $V \sim 3 \Mp^2 H^2$), with the observable level fortuitously corresponding to GUT scale inflation. 
However, since $V^{1/4}\sim (r/0.01)^{1/4}\times 10^{16}$ GeV, improvements in $r$ of the order we consider in this document do not translate into a large improvement in the constraint on the scale $V^{1/4}$.   However, the improvements do correspond to reaching a very significant threshold in the field range of the inflaton as described next.

\smallskip

\noindent (3)  Either a detection or a constraint at the level of $r\sim 0.01 - 0.001$ would determine whether the inflaton $\phi$ rolled a super-Planckian or sub-Planckian distance in field space.  In single field slow-roll inflation, we have the following simple relation \cite{Lyth:1996im} relating the number of e-foldings $N_e$ to the tensor to scalar ratio $r$
\beq\label{Lyth}
N_e=\int \frac{da}{a}=\int H dt = \int \frac{H \Mp }{\dot\phi}\frac{d\phi}{\Mp} = \sqrt{8} \,  r^{-1/2}\frac{\Delta\phi}{\Mp}
\eeq
where we used the slow-roll result $\langle\zeta\zeta\rangle' = \frac{H^4}{2\dot\phi^2}$, (\ref{tensortwo}), and (\ref{rthreshold}).  Again, we have assumed that the background inflationary solution is evolving slowly, enabling us to take $r$ out of the integral.    
A detection of $r$ corresponding to $\Delta\phi\ge \Mp $ would imply sensitivity to an infinite sequence of Planck-suppressed operators in the effective action for the inflaton.  At a field range of $M_P$, a minimal reheat temperature roughly corresponds to $N_e\sim 30$ and $r\sim 0.01$, and a maximal reheating temperature corresponds to $N_e\sim 60$ and $r\sim 0.002$.  The former is the reach of near-term experiments, and the latter the goal of the Stage IV experiment, \cmbexp, as discussed 
in the experimental sections of this report. 

An approximate shift symmetry---if present in the ultraviolet completion of 
gravity---may protect against such terms.  Such a symmetry over a large field range yields radiative stability from the low-energy quantum field theory point of view.  Two now-classic examples are $m^2\phi^2$ inflation~\cite{Linde:1983gd} and Natural Inflation~\cite{Freese:1990rb} (traditional quantum field theory axions, requiring a super-Planckian period), as well as a generalization known as N-flation~\cite{Dimopoulos:2005ac} which employs multiple fields each with a small range.  
A recent theoretical development is the recognition of a very broad class of different models, including the Starobinsky model
\cite{Starobinsky:1980te}, Higgs inflation \cite{Salopek:1988qh,Ellis:2013xoa,Bezrukov:2013fca}, and a broad class of models based on spontaneously broken conformal and superconformal symmetry \cite{Kallosh:2013hoa}, which give identical, model-independent predictions of $n_s$ and $r$, with the caveat that these models still await ultraviolet completion. In particular, these models predict tensor perturbations with $r \sim 0.004$, within reach of the experimental program described in this report.

It is important to understand if a symmetry over a super-Planckian range operates in a consistent ultraviolet completion of gravity.  In string theory, a leading candidate for quantum gravity, a way in which large-field inflation protected by an underlying shift symmetry arises naturally is via monodromy~\cite{Silverstein:2008sg, McAllister:2008hb, Kaloper:2011jz} for scalar fields such as axions:  in the presence of generic fluxes and branes, potential energy builds up along axion directions.  This yields potentials which flatten out relative to $m^2\phi^2$ at large field range, a simple result of massive degrees of freedom adjusting in an energetically favorable way \cite{flattening}.   This mechanism predicts observable $B$ modes and a tilt distinct from $m^2\phi^2$ inflation and Natural Inflation.  
Other string theoretic mechanisms can produce small field inflation, a perfectly viable possibility as well.      

Another interesting direction is to use tensor perturbations to constrain additional sectors of light fields.  For various scenarios involving additional axions, this has been extensively explored in the literature; see for example \cite{Hertzberg:2008wr}\cite{Kaplan:2008ss}\ and references therein.  

\smallskip

To summarize, the CMB $B$-mode measurements will, according to present forecasts, reach an important threshold, the Planck range in field.  A null result on $r$ would be qualitatively important, showing that small field inflation is the operative case.   A detection would provide an important lever to quantum gravitational physics.  

\smallskip

On the experimental side, the Stage-IV CMB experiment \cmbexp\ described below is designed to make
major improvements in sensitivity to tensor fluctuations over current experiments
and {\em provide a {conclusive} experimental discrimination between large-field and small-field inflation models}.
The proposed goal for \cmbexp\ on the precision of the tensor to scalar ratio $r$ is ${\bf \sigma(r)=0.001}$ (statistical). 
Although this goal is ambitious, it is found to be achievable even over a wide range of potential \cmbexp\ survey
designs, using only the $\ell\sim 100$ information that is readily accessible to ground-based instruments.
With this level of uncertainty,
\cmbexp\ will make a clear {\em detection} ($> 5\sigma$, as universally accepted in the HEP community) of tensor modes from any large-field inflation model ($r\gtrsim 0.01$).  If $r$ is near the $2\sigma$ limit set by data from the \planck\ satellite---as predicted by $m^2\phi^2$ inflation---primordial $B$-mode polarization is within the detection reach of ongoing and currently funded experiments.  Under this scenario, the \cmbexp\ surveys can be reconfigured to measure the tensor amplitude with high precision and to characterize the more detailed properties of the tensor fluctuations, such as their scale-invariance and Gaussianity or lack thereof. These measurements, even with relatively modest precision, would constitute striking qualitative tests of inflation models.

\begin{figure}[t]
\begin{center}
\includegraphics[width=0.7\textwidth]{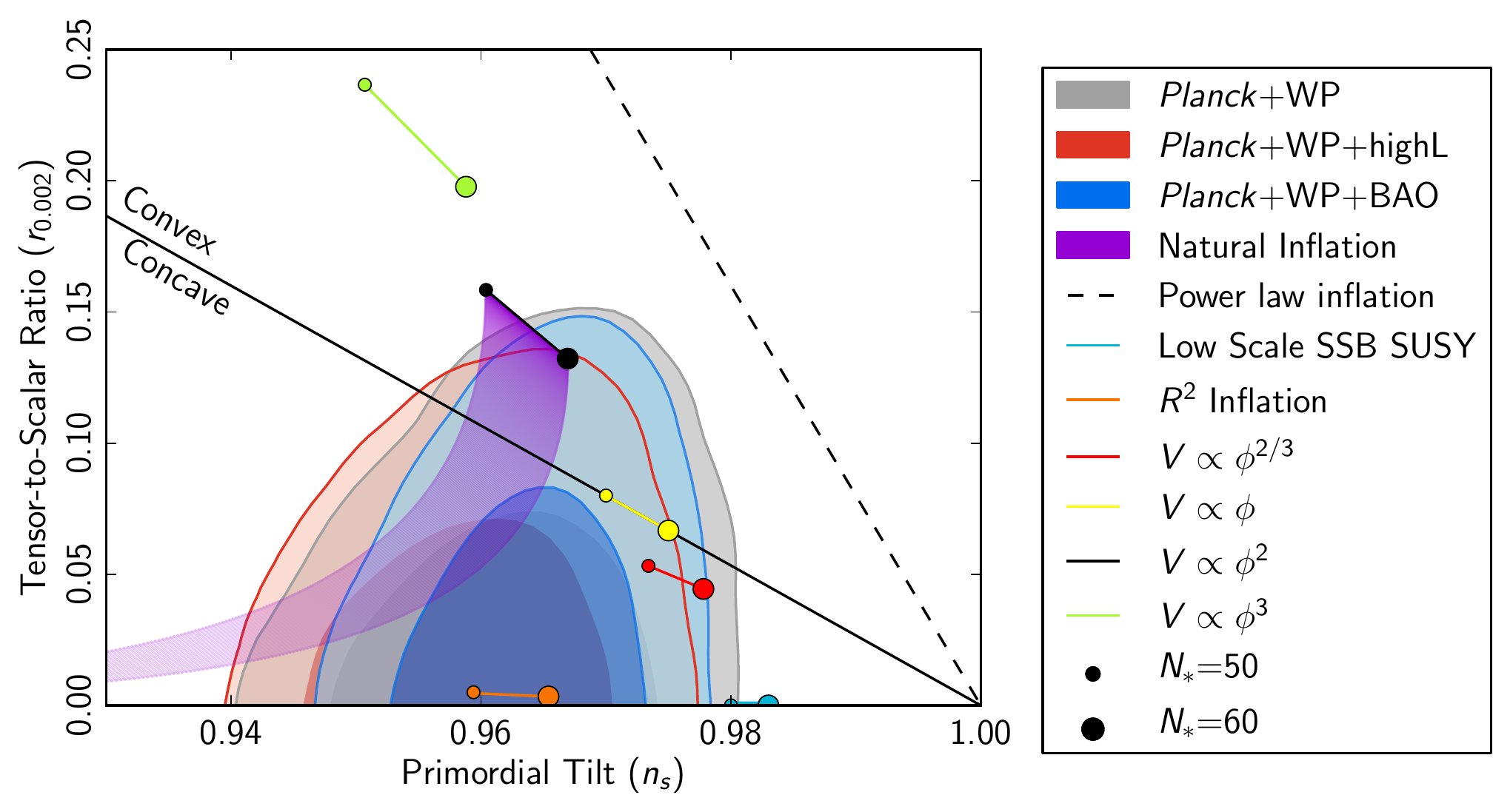}
\caption{\label{fig:nsr} \small\it Current CMB constraints on the
  combined $n_s$-$r$ parameter space \cite{Ade:2013uln}. Reproduced with permission from Astronomy \& Astrophysics, copyright ESO; original
source ESA and the Planck Collaboration.}
\end{center}
\end{figure}

\subsection{Scalar Tilt, Running, and Beyond}
During inflation $|\dot H |, |\ddot H| >0$ and therefore the background evolves with time.  Perturbations produced at different times (and hence with different $k$) see different Hubble scales and will freeze with slightly different amplitudes.  As a result, most inflation models predict a power law dependence of the scalar power spectrum
\beq
 k^3 \langle \zeta_{\k} \zeta_{-\k} \rangle   \propto     k^{n_s-1} 
\eeq
with a value of $n_s$ slightly different from the value $n_s =1 $ in the scale-invariant case.  Current measurements from 
the \planck\  satellite give  ~\cite{Ade:2013uln}
\beq
   n_s  = 0.960 \pm 0.0073
\eeq
excluding $n_s=1$ by more than $5\sigma$.  This is concrete evidence that cosmic structure
 was generated by a physical quantum
field theory rather than by any putative mechanism imposing precise scale-invariance. 

In the slow-roll approximation, the exponent $n_s$ is given by \beq
n_s - 1 = - 6 \epsilon + 2 \eta \eeq where \beq \epsilon =
\frac{\Mp^2}{2} \left(\frac{V'}{V} \right)^2 \qquad \eta = \Mp^2
\frac{V''}{V} \ , \eeq with $ ' \equiv \tfrac{\partial}{\partial
  \phi}$.  Thus, the precision measurement of $n_s$ gives specific
information that can be used as a constraint on a model of the
inflaton potential.  This is especially relevant when combined with \planck\ 
measurements of $r$.  From Figure~\ref{fig:nsr}, we can see that
various classes of large-field models make distinct predictions within
the $n_s$-$r$ plane.  Of particular interest is the fate of the
minimal $m^2 \phi^2$ model, which is on the edge of the currently
allowed region.  Therefore, we gain critical information through $n_s$
and its running, $\frac{\partial n_s}{\partial \log k}$, that is
complementary to the effort to measure $r$.

Finally, inflation predicts small, but non-zero curvature.  In a
general inflationary scenario, one expects spatial curvature on the
horizon patches to receive contributions from large scale modes and
hence the curvature parameter $\Omega_k$  is naturally
of the order $10^{-4}$. Measurements of the value of $\Omega_k$ that
would differ from naive inflationary expectation would be very
informative about the process responsible for inflation.

In particular, if $|\Omega_k|$ is found to be considerably larger than
this value, it will tell us that inflation was not proceeding in a
slow-roll when scales just larger than our observable horizon exited
their inflationary horizons. More concretely, observations of negative
and large $\Omega_k$ (positive curvature) would falsify 
eternal inflation, 
while observation of positive and large $\Omega_k$
(negative curvature) would be consistent with false vacuum eternal inflation
\cite{Kleban:2012ph,Guth:2012ww}.

Our current constraints on this parameter from the CMB alone are
$\Omega_k=0.042^{+0.027}_{-0.018}$ and improve significantly to
$\Omega_k= (0.5\pm3.3) \times 10^{-3}$ upon the addition of baryon
acoustic oscillation (BAO) data \cite{Ade:2013lta}. While these
numbers are already interesting, future constraints approach an even
more interesting regime. For example, an optimistic forecast for the
error on $\Omega_k$ stemming from combination of current \planck\
priors, \textit{LSST} and Dark Energy Spectroscopic Instrument (DESI)
data (assuming $\Lambda$CDM, but varying neutrino masses) is $9\times
10^{-4}$, significantly improving current constraints
\cite{Font-Ribera:2013rwa}. If curvature were to be detected at this
level, it would have profound implications for the inflationary
paradigm.

\smallskip

Further potential observables include features in the potential or oscillations in the power spectrum (see e.g., \cite{Starobinsky:1998mj, Easther:2001fi, Flauger:2009ab,  Achucarro:2010da, Chen:2011zf}) which may be correlated with measurable $B$ modes \cite{Flauger:2009ab}.  The latter analysis is subtle because of theoretical backreaction effects, offering an interesting challenge at the interface of theory and data analysis.

\smallskip

Finally, anomalies reported by the \planck\  collaboration, which would also require additional parameters beyond $\Lambda$CDM, deserve further work.   These may indicate additional structure in the power spectrum and non-Gaussianity, perhaps related to pre-inflationary relics or reheating dynamics in the presence of additional fields.  However, their statistical significance is currently low. The wavenumber of the modes which contribute the most to the multipoles of the anomalies is not particularly small. Next generation LSS surveys, such as the Large Synoptic Survey Telescope (LSST) and DESI, will be able to observe some of these modes, although it is still unclear if they will be able to control systematics effects at such large distances. Since for a three-dimensional survey error bars from cosmic variance are inversely proportional to the wavenumber of the mode to the 3/2 power, there is, in principle, enough information to make the \planck\  anomalies significant by many standard deviations. Clearly, this would be a very important discovery that requires further study.

\subsection{Non-Gaussianity}
\label{sec:png}
More direct evidence of the nature of the inflaton as a physical interacting quantum field would come from the observation
of nonlinear inflaton interactions.  These would show up as correlations in the CMB fluctuations beyond the Gaussian
approximation (see \cite{Chen:2010xka} for a recent review).  The presence of non-Gaussian correlations in the
initial condition for structure formation should also be visible in
direct observations of cosmic structure, measured by large scale
surveys such as LSST and DESI.  They would appear as non-Gaussian correlations of galaxies (or dark matter) on very large scales.

Any observation of a non-zero correlation function beyond the power spectra (i.e., non-Gaussianity) would include a wealth of new information about the primordial universe.  Many studies have focused on a non-zero three point function of primordial perturbations, which can be expressed in terms of the bispectrum $B(k_1,k_2,k_3)$,
\beq\label{bispectrum}
\langle \zeta_{\k_1}\zeta_{\k_2}\zeta_{\k_3}\rangle = B(k_1,k_2,k_3) \ (2\pi)^3 \delta^{(3)}(\k_1+\k_2+\k_3)\ .
\eeq
The four-point function has been explored as well, though to a lesser extent.
After considering translation and rotation invariance (as assumed in (\ref{bispectrum})), the bispectrum is a function of three variables.  Many models produce scale invariant correlation functions, which further reduces it to a function of two variables that can be described in terms of the shape of the triangle in momentum space formed from $\k_1,\k_2,\k_3$ \cite{Babich:2004gb}.  Large violations of scale invariance can arise in the bispectrum (e.g., \cite{Chen:2006xjb, Flauger:2010ja}) and may \cite{Behbahani:2011it} (or may not \cite{Behbahani:2012be}) have correlated signatures in the power spectrum.

These higher correlation functions contain full functions worth of information about the inflationary epoch.  This information would be invaluable to our understanding of the mechanism of inflation and the origin of the primordial fluctuations.  Some important implications include:
\vskip 6pt
(1)  {\it Probing interactions:}  Inflation requires a slowly varying Hubble expansion, for example a slow variation in the potential energy of a field.  This can arise roughly speaking in two ways---either via a flat potential, or on a steep potential with sufficient interactions or dissipation to slow the field evolution.  The latter class of mechanisms tends to produce significant non-Gaussianity, peaked on equilateral triangles $k_1\sim k_2\sim k_3$ (with other shapes possible).   UV-complete mechanisms exhibiting all these possibilities have been studied \cite{Silverstein:2003hf, Alishahiha:2004eh},  helping stimulate the much more systematic understanding of the observables \cite{Chen:2006nt} that have been classified using the effective field theory (EFT) of the inflationary perturbations \cite{Creminelli:2006xe, Cheung:2007st}.

In the EFT, the inflationary perturbations can be realized as Goldstone modes ($\pi$) of time translation symmetry, whose dynamics is described by a universal Lagrangian, schematically
\begin{eqnarray}\label{Spi} 
\! \! S_{\rm \pi} = \int \!  d^4 x   \sqrt{- g} \left[M_P^2
\dot{H} \left(\dot\pi^2-\frac{ (\partial_i \pi)^2}{a^2}\right)
+ M^4_2
\left(\dot\pi^2+\dot{\pi}^3-\dot\pi\frac{(\partial_i\pi)^2}{a^2}
\right) + M^4_3 \dot{\pi}^3+ ... \right],
\end{eqnarray}
where $\pi$ is the Goldstone boson associated to the non-linearly realized time-translations. 
As typical of Goldstone bosons, this Lagrangian is characterized by higher-dimension interactions that violate perturbative unitarity at an high energy scale $\Lambda_U$~\cite{Cheung:2007st, Baumann:2011su}. Similarly to what happens in particle physics with the precision electroweak tests, limits on non-Gaussianities can be mapped, in a model-independent way, into limits of the parameters of this Lagrangian, as recently done by the WMAP and \planck\  collaborations. The amplitude on non-Gaussianities take the form $(H / \Lambda_U)^{\alpha}$ where $\alpha \sim 2$.  In the case of single field inflation, existing constraints from \planck\  on the equilateral and the ``orthogonal" shapes require $\Lambda_U \gtrsim {\cal O}(10) H$. It would be extremely interesting to raise the bounds on $\Lambda_U$ in order to constrain it to be larger than the scale associated to the speed of inflaton in standard slow-roll models, 
given by $\dot\phi^{1/2}\sim {\cal{O}}(100) H$ (or equivalently $f_{\rm NL}^{\rm equilateral} \sim 1$). 
In this case, we would have shown that the EFT can be extended to such a high energy to be best described by standard slow-roll inflation.

\vskip 6pt
(2)  {\it ``Discovering" extra fields:} It can be shown that single-field inflation produces no signal in the ``squeezed limit" of the bispectrum where one momentum is much smaller than the other two, $k_1\ll k_2\sim k_3$ \cite{Maldacena:2002vr, Creminelli:2004yq}.  The constraints on this limit are often characterized in terms of the local shape, $\fnl^{\rm local}$.  The detection of a non-zero $\fnl^{\rm local}$, or any non-trivial squeezed shape, would require additional fields, beyond the inflaton, and thus would rule out single-field inflation.  Furthermore, precise measurements in the limit can uncover non-trivial information regarding these extra fields.  For example, additional sectors containing weakly interacting massive fields~\cite{Chen:2009zp} (as in supersymmetric theories~\cite{Baumann:2011nk}), as well as additional sectors of strongly coupled CFTs~\cite{Green:2013rd},  both produce bispectra of the form $B_{{\rm squeezed}}(k_1,k_2,k_3)(k_1/k_2)^\Delta$ (where the constant $\Delta$ is determined by the mass, via $\Delta = \tfrac{3}{2} - \sqrt{\tfrac{9}{4} -\tfrac{m^2}{H^2}}$, or is the  dimension of a conformal primary operator).  The measurement of $\Delta$ would thus tell us about the spectrum of masses / operators in such a sector.

Given the \planck\  constraint on $\fnl^{\rm local}$, it is possible to turn things around and derive precision constraints on hidden sectors coupled to the inflaton perturbations via higher dimension operators suppressed by a mass scale $M_*$; this scale must be several orders of magnitude larger than the Hubble scale during inflation \cite{Green:2013rd, Assassi:2013gxa}.  For large-field inflation that provides a lever to the Planck scale.

Finally, there are several additional forms of non-Gaussianities that are not characterized by a peculiar squeezed limit, that are largely unconstrained, and that are possible only in multifield models. They have emerged in the context of the effective field theory approach to multi field inflation~\cite{Senatore:2010wk}, and they can be mapped into signatures of the symmetries that explain the lightness of  additional fields  during inflation. 
\vskip 6pt
(3)  {\it ``Discovering" multiple sources:}
Relations between the bispectrum and the trispectrum (4-point function) can further indicate that multiple fields produce the observed curvature perturbations (i.e., there are multiple sources of curvature).  Specifically, if the amplitude of the four point function, $\tau_{\rm NL}$, is found to satisfy $\tau_{\rm NL} \gg f_{\rm NL}^2$, then one must have multiple sources of the curvature perturbation \cite{Suyama:2010uj, Barnaby:2011pe, Byrnes:2011ri, Assassi:2012zq}.  Furthermore, the consistency of many models requires that the precise shape of the bispectrum and trispectrum are not independent \cite{Cheung:2007st}.  Precise measurements of both would then strongly constrain the origin of the curvature perturbations.

\smallskip

From all this it is clear that an improved determination of the constraints on non-Gaussianity, or a detection, would help distinguish very different mechanisms for inflation and possibilities for additional sectors in the early universe. 
Moreover, it is worth stressing that other forms of non-Gaussianity, not yet searched for, may lie in the data.  

\smallskip

To date, the best constraints on non-Gaussianity have come from
observations of the CMB.  However, current and planned large scale
structure surveys like the Baryon Oscillation Spectroscopic Survey
(BOSS) \cite{Dawson:2012va}, its successor eBOSS, DESI, Dark Energy
Survey (DES) and LSST have the potential to dramatically improve our
sensitivity to non-Gaussian effects beyond the limits set by \planck.
Realizing the promise of these surveys for inflation requires a
community effort dedicated to both data analysis for non-Gaussianity
and improving our theoretical understanding of the mildly non-linear
regime of large scale structure (including dark-matter clustering,
tracer biasing and redshift-space distortions).  These efforts will be
relatively inexpensive, and their value could be enormous.  Since the
amount of information grows like the cube of the smallest scale
measured, a factor of two improvement in our understanding of the
mildly non-linear regime is equivalent to improving the
size of a survey almost tenfold.  A recently developed approach, based on applying
effective field theory techniques to LSS~\cite{Carrasco:2012cv}, seem
to show that this is indeed possible.  Such efforts would pay similar
dividends to the study of dark energy and are therefore complementary
to the existing effort.

\section{Constraining Inflation Physics With Cosmological Probes}
\label{sec:measintro}
As detailed in Section~\ref{sec:theory}, the theory of cosmic inflation
is the most promising model for the dynamics of the universe at very early times
and high energies. 
Inflation supplies a natural explanation
for the smoothness 
and geometrical flatness 
of our observable universe, and the predicted consequences of an inflationary
beginning to our universe, including a nearly Gaussian distribution of primarily adiabatic density 
fluctuations with a nearly scale-invariant spectrum, have been spectacularly confirmed in recent
years. The strong acoustic features in the power spectrum of fluctuations in the 
CMB temperature and polarization fields and the correlations between the 
temperature and polarization fluctuations demonstrate that the primordial density fluctuations
are almost entirely adiabatic \cite{Larson:2010gs}.  The power-law slope of the primordial
fluctuations has recently been shown to be very close to the scale-invariant value but with
a significant ($5 \sigma$) deviation in the direction predicted by inflation \cite{Ade:2013lta}. 
The measured level of non-Gaussianity in the CMB has remained consistent with zero, even as 
experimental sensitivity has dramatically increased \cite{Komatsu:2003fd,Ade:2013ydc}.

Now that the general picture of an early inflationary epoch is well-established, we 
can begin to ask questions about the specifics of inflation physics, such as: What was the 
energy scale of inflation? Was inflation caused by a single scalar field, or were there multiple
fields involved? What was the shape of the inflationary potential? The next generation of 
cosmological probes, particularly measurements of the polarization of the CMB and
large-area galaxy surveys, are poised to deliver the first answers to these fundamental 
questions.

\subsection{Probing High Energy Scale Physics with the Polarized Cosmic Microwave Background}
\label{bmodeintro}
The study of the CMB is a unique and powerful tool for learning how our universe works
at the most fundamental level.  Ever since the epoch of recombination, when the CMB was released,
most of the photons in the universe have been freely streaming.  We can therefore use the CMB to directly
test precise predictions of cosmological models and directly
probe the basic physics governing our universe up to the epoch of recombination, including 
models of cosmic inflation in the very early universe.

One of the inescapable predictions of inflation is that the period of violent 
expansion produced a spectrum of relic gravitational waves (i.e., tensor-mode perturbations).
The ratio of the gravitational-wave (tensor) to density-fluctuation 
(scalar) power in the CMB, $r$, 
is related to the energy scale of inflation and the range of the inflaton field as discussed above in
Section \ref{sec:theory}.
Even though these tensor modes are produced at all wavelengths, the signal 
is far too small to be seen directly with any conceivable gravitational-wave interferometer.
However, these gravitational waves leave a potentially observable 
imprint in the CMB at large angular scales. 
If the energy scale of inflation were high enough, the primordial gravitational-wave background 
would have been seen by the WMAP or \planck\  satellites via the extra fluctuations that 
would have been left behind in the CMB temperature power spectrum.
\planck\  has placed a $95\%$ upper limit on the tensor-to-scalar ratio of
$r < 0.11$~\cite{Ade:2013lta}.  Due to cosmic variance (the fact that we only have
one universe to observe), we will never be able to improve this limit on $r$ significantly through
CMB temperature measurements alone.

This leaves us with only one avenue to probe the energy scale of inflation: precision measurements
of the polarization of the CMB.
Primordial gravitational waves generated as a result of inflation would have imprinted a faint but unique
signal in the polarization of the CMB, the amplitude of which directly scales with the energy scale of
inflation. The polarization of the CMB can be decomposed
into a curl-free component and a divergence-free component, called $E$ modes and $B$ modes
in analogy to electromagnetism. The polarization imprinted on the CMB at the last scattering
surface by a scalar-generated quadrupole can only be an $E$-mode signal 
and has been well-measured~\cite{brown09, chiang10, quiet11,Bennett:2012zja}.  In contrast, the only expected source of $B$-mode signal 
in the primary CMB is gravitational waves from inflation (active sources after inflation such as cosmic strings
and defects from phase transitions could contribute, but the angular dependence of the $B$ modes
from these sources should make them clearly distinguishable from inflationary $B$ modes).  
If inflation happened on the 
GUT scale ($\sim 10^{16}$~GeV), the signal will be easily detectable, and with a future generation of CMB
experiments we will be able to directly map the quantum fluctuations generated by inflation.

A secondary signal from gravitational lensing of the CMB by structure along the line of sight
can produce $B$ modes by distorting modes that were 
originally pure $E$.
This lensing $B$-mode signal
is predominantly at smaller angular scales than the inflationary signal and will have to be well-measured
and cleaned (or ``delensed'') from CMB maps in order to continue pushing our ability to measure the primordial
gravitational-wave signal. The lensing $B$-mode signal is itself an exciting cosmological probe,
and there is compelling independent particle physics motivation to measure these modes well 
(see the report {\it Neutrino Physics from the Cosmic Microwave Background and Large Scale Structure} \cite{Abazajian:2013oma}).
Alternatively, the lensing $B$-mode power spectrum can be predicted from theory and subtracted from the 
measured power spectrum (``debiasing'').

\begin{figure}[t!]
\begin{center}
\includegraphics[width=0.9\hsize]{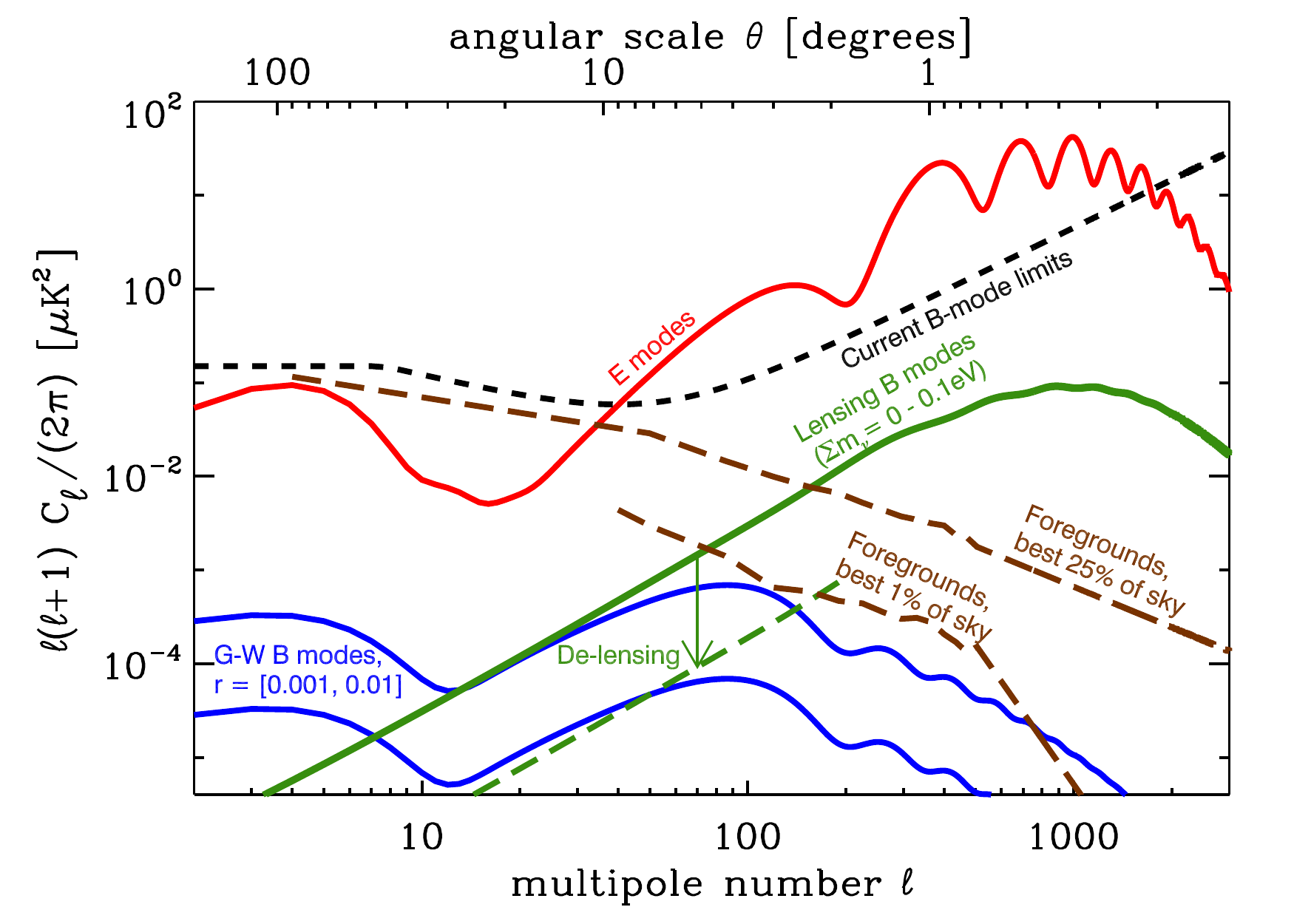}
\caption{Expected signal levels for the $E$-mode (\textbf{red, solid}), 
inflationary gravitational-wave $B$-mode (\textbf{blue, solid}),
and lensing $B$-mode  (\textbf{green, solid}) signals.
The gravitational-wave $B$-mode signals are shown for tensor-to-scalar 
ratios of $r=0.001$ (the Stage-IV goal) and $r=0.01$ (the boundary between 
small-field and large-field inflation models). The 
lensing $B$-mode signal is shown as a band encompassing the predicted signal for 
values of the sum of neutrino masses $0 \le \sum m_\nu \le 0.1 \mathrm{eV}$. 
Delensing by a factor of 4 in amplitude is shown schematically by the \textbf{green arrow},
with the residual signal at $\ell \le 200$ (where the delensing is critical to the constraint 
on $r$) shown by the \textbf{green, long-dashed} line.
The \textbf{black, short-dashed} line shows the level of current 95\% upper 
limits on $B$ modes from \textsc{WMAP} at the largest scales, the \textsc{BICEP} experiment at degree scales, 
and the \textsc{QUIET} and \textsc{QUaD} experiments at smaller scales. The 
\textbf{brown, long-dashed} lines show the expected 
polarized foreground contamination at 95~GHz for the cleanest $1\%$ and $25\%$ of the sky.}
\label{fig:powspecintro}
\end{center}
\end{figure}

Figure \ref{fig:powspecintro} shows the expected signal levels for the scalar $E$-mode, 
inflationary gravitational-wave $B$-mode, and lensing $B$-mode signals (with and without 
delensing by a factor of four in amplitude or 16 in power) for tensor-to-scalar ratios of $r=0.001$ and $r=0.01.$ Also shown 
are current limits on the $B$-mode amplitude and expected foreground contamination
at an observing frequency of 95~GHz for the cleanest $1\%$ and $25\%$ of the sky.
As Figure \ref{fig:powspecintro} shows, the inflationary $B$-mode signal is accessible, particularly near
the $\ell = 100$ peak, but a detection of $r \sim 0.01$ will require significant improvements in instrument sensitivity and 
careful control of foreground and lensing contamination. 

Sections \ref{sec:currentexp} through \ref{sec:survey} provide an overview of the current generation
of CMB polarization experiments and the inflationary science reach of a potential future experiment.
We have adopted a nomenclature to categorize
these current and future efforts: we categorize generations, 
or stages, of experiments roughly by the number of detectors that are on the sky, which is a good 
proxy for sensitivity.  Stage-I experiments have finished observing, and have roughly $\sim 100$ detectors.  
Stage-II experiments are currently observing with $~\sim 1000$ detectors, and Stage-III experiments are currently 
under development with $~\sim 10000$ detectors.  A Stage-IV experiment would have yet another order of magnitude 
more detectors and would harness the resources and experience of the CMB community to produce a cohesive suite 
of experiments targeting some of the most interesting and fundamental questions in the study of the 
nature of our universe.

\begin{figure}[tbh]
\begin{center}
\includegraphics[width=0.9\hsize]{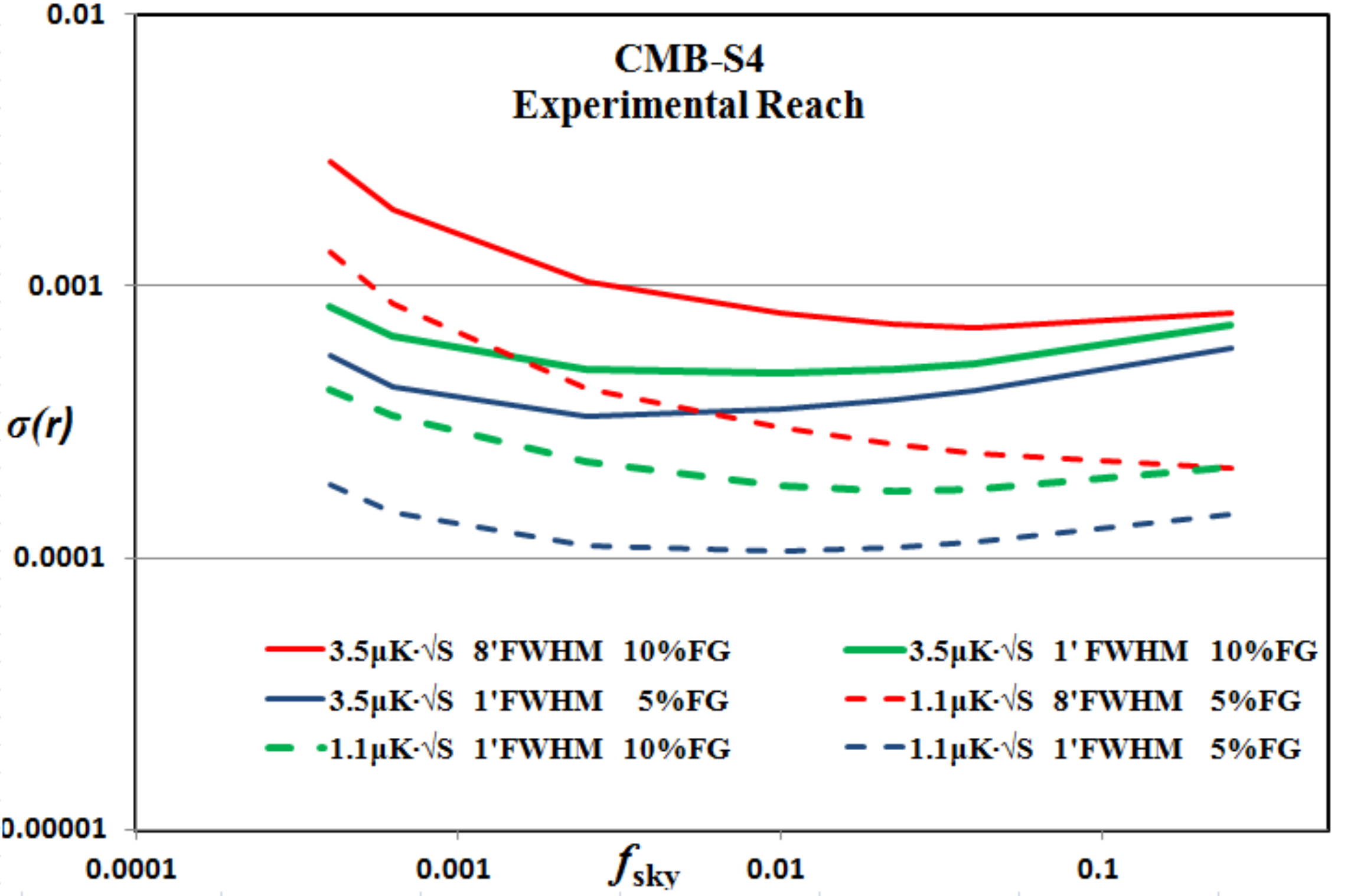}
\caption{ Expected statistical uncertainty on the tensor-to-scalar ratio $r$ for a few Stage-IV CMB 
configurations, plotted as a function of sky coverage.  Forecasts for two beam sizes 
($1'$ and $8'$ FWHM) and two overall sensitivity goals ($3.5$ and $1.1$ $\muk \sqrt{s}$) are
displayed.  Foregrounds are assumed to be cleaned to a residual level of $10\%$ or $5\%$ (in amplitude).  We find 
the stated goal of $\sigma(r)<10^{-3}$ is achieved or exceeded.}
\label{fig:tensor}
\end{center}
\end{figure}

Figure \ref{fig:tensor} shows the results of forecasts for a few representative configurations of a Stage-IV CMB 
experiment (for details, see Section \ref{sec:survey}).  The Stage-IV goal of $\sigma(r)=10^{-3}$---which would
result in an unambiguous $>5 \sigma$ detection for large-field inflation---is achieved or 
exceeded for a variety of design parameters, fractions of sky covered ($\fsky$), and foreground assumptions, 
with a broad minimum around $\fsky=1\%$.  

\subsubsection{Current Experimental Efforts and Upgrades: Stages II and III}
\label{sec:currentexp}
The current generation of CMB polarization experiments (Stage II) consists of a suite of complementary 
experiments that employ a variety of experimental approaches to take steps toward searching for an 
inflationary $B$-mode signal.  There are a number of experiments that observe the sky on degree angular 
scales, and are specifically designed to directly measure inflationary $B$ modes.  If the level
of tensor modes is near the upper limit of what is still allowed by recent \planck\  results~\cite{Ade:2013lta},
these Stage II experiments have the sensitivity required to measure a $B$-mode signal.

The \textsc{BICEP/Keck Array}~\cite{nguyen08} program at the South Pole is a set of 
compact, refractive polarimeters currently taking data on the sky at degree angular scales
and has already accumulated high raw sensitivity at $150$~GHz.  The \textsc{BICEP} program will 
expand its frequency coverage to include $100$~GHz and $220$~GHz, adding a Stage-III experiment
in 2015.  The balloon borne \textsc{EBEX}~\cite{reichborn-kjennerud2010} experiment has already 
collected data over 6000 square degrees with three frequency bands between 150 and 410 GHz in a flight 
launched in December 2012. \textsc{SPIDER} ~\cite{Fraisse2013} and \textsc{PIPER}, also balloon-borne, will 
have first light at the end of 2013 and 2014, respectively, and will observe large areas of the sky at 
frequencies between 100 and 600~GHz.  Although the integration time of balloon-borne 
experiments is much shorter than ground-based experiments, they benefit from dramatically reduced
atmospheric contamination and detector loading. 
The \textsc{ABS}~\cite{essinger10} experiment observes in Chile at degree angular scales and \textsc{CLASS} will begin 
observations in 2014. 
All of these experiments employ different and complementary approaches to mitigation of systematic uncertainties.

Because of their large beams and limited frequency coverage, these degree-angular-scale 
Stage-II and Stage-III programs are expected to be limited ultimately by foregrounds and
the gravitational lensing signal.  Pushing harder on inflationary science 
requires input from experiments with higher angular 
resolution and broader frequency coverage to disentangle a primordial $B$-mode signal from a 
foreground or gravitational lensing signal.  Joint analysis with higher angular resolution data 
from Stage-II and Stage-III experiments such as \textsc{POLARBEAR}~\cite{Kermish:2012eh}, 
\textsc{SPTpol}~\cite{Austermann:2012ga}, and \textsc{ACTpol}~\cite{Niemack:2010wz} 
will become critical in the near future 
when lensing becomes the limiting factor in the search for the inflationary signature.
Partnerships have already formed between complementary 
experiments to begin joint analysis of data sets for delensing.
Of these experiments that have sensitivity on arcminute angular scales, \textsc{POLARBEAR} and \textsc{SPTpol}
have been on-sky for over a year---\textsc{SPTpol} recently reported the first detection of lensing $B$ modes \cite{Hanson:2013daa}---and \textsc{ACTpol} is currenly being deployed.  Stage-III upgrades
for all three of these experiments are planned in the next few years.
 
\subsubsection{The Stage-IV Experimental Goal on the Tensor-to-Scalar Ratio: ${\bf \sigma(r)=0.001}$.}
\label{sec:design}
The Stage-IV CMB experiment aims to significantly improve sensitivity over current efforts
and to provide a conclusive experimental discrimination of large-field and small-field inflation.  
The proposed sensitivity goal for a Stage-IV CMB experiment is to measure the tensor-to-scalar 
ratio to ${\bf \sigma(r)=0.001}$ (stat.), with a similar level of systematic
uncertainty budgeted. 
With this level of uncertainty, 
a Stage-IV CMB experiment will unambiguously detect (CL $> 5\sigma$) tensor modes from any large-field inflation 
model with $r \gtrsim 0.01$.  Conversely, a null result at the level of $r< 0.002$ would rule out large-field inflation.
A Stage-IV experiment will determine if inflation happened on the GUT scale or not.
This goal is achievable with a wide range of Stage-IV design parameters
targeted at measuring the degree-scale feature in the $B$-mode spectrum 
($\ell\sim 100$) known to be accessible from the ground.

If $r$ is near the $2\sigma$ \planck\  limit, as in the case of $m^2\phi^2$ inflation, current funded 
Stage-III experiments will detect $B$-mode polarization with high significance.  
In that scenario, a Stage-IV survey can be designed to measure the tensor amplitude with high 
precision and directly map quantum fluctuations produced during inflation.  
Further characterization of tensor properties, such as scale-invariance and level of non-Gaussianity, 
will be possible with Stage IV's superb sensitivity.  
For instance, a survey with $\sim$$1 \; \muk$-arcmin noise over roughly half the sky would 
be able to make a measurement of the the tensor spectral index $n_t$ at a signal-to-noise
of $\sim$2, assuming the slow-roll consistency relation $n_t = -r/8$ holds; 
put another way, large deviations from this consistency relation in the blue direction would
be strongly ruled out.
Tests such as these, even with relatively modest 
precision, go beyond simple verification of inflation to detailed discrimination between 
inflation models.

\subsubsection{Survey Considerations for a Stage-IV experiment}
\label{sec:survey}
Achieving the design goals of the Stage-IV experiment outlined in Section~\ref{sec:design}
will require a major advancement in raw sensitivity and tight control of instrumental
systematics. 
Measuring the tensor-to-scalar
ratio $r$ below a level of $\sigma(r)=0.01$ requires mapping at least hundreds of 
square degrees of sky to noise levels well below $10 \ \muk$ in a 1-arcmin pixel 
(or $10 \ \muk$-arcmin), which is the lowest noise level achieved in any current CMB 
observation of a patch of sky larger than a few square degrees. 
Effectively cleaning
the lensing $B$ modes from the inflationary signal also places stringent requirements on sensitivity. 
Significant delensing is only possible with a $< 10^\prime$ 
beam and $< 5 \ \muk$-arcmin noise in the $B$-mode map~\cite{Seljak:2003pn}.  This noise level and resolution across
a significant area of sky (at least $1000 \ \sqdeg$) would reduce the lensing limit on the
measurement of $r$ to below $10^{-3}$. 

A Stage-IV polarized CMB experiment must be designed with a few key features in mind.
It must have multiple observing bands to separate polarized foreground signals.  It must 
have the sensitivity required on degree angular scales to observe the primordial signature
and on arcminute angular scales for effective delensing.  It must maintain a clean instrumental 
and optical design to reduce spurious polarized signals.  A balance must be found
between the size of the observed sky region and the depth to which that region is mapped.

The size of the observed patch is determined by the 
angular scale of the primordial $B$-mode feature accessible to ground based instruments, 
the so-called ``recombination bump'' at multipole $\ell \sim 100$, or angular
scale $\theta \sim 2^\circ$. According to theoretical predictions, the inflationary 
tensor signal at very large angular scales, the ``reionization bump''
at $\ell <10$, exceeds lensing even for low values of $r$ ($r<10^{-3}$).  
Predicted galactic dust foreground at these large angular scales is expected to be a factor of 30
(in temperature) larger than the signal for $r=0.001$ (at 90~GHz, for 5\% dust polarization 
and for 75\% of the sky~\cite{dunkley09}). 
Information from \planck\  will provide more information 
about polarized foregrounds.  In this document we are describing a 
\cmbexp\ experiment that will extract the primordial $B$-mode signal from a deep survey over only a 
few percent of relatively clean sky. 

Lensing-induced $B$ modes have the same frequency dependence as the CMB and cannot be distinguished by 
multi-frequency observations. 
The lensing signal peaks at $\ell\sim 1000$ and tends to increase the optimal survey width,
because covering more sky allows lensing confusion to be debiased (i.e., subtracted in power-spectrum space) in the same 
way instrumental noise is removed in temperature power spectrum measurements.  This is possible because the lensing 
amplitude is determined to within $3\%$ by constraints on cosmological parameters from \planck's temperature 
and $E$-mode measurements, and the uncertainty is expected to decrease dramatically, using the arcminute-scale 
$B$-mode survey enabled by a Stage-IV experiment.  
Alternatively, the lensing deflection field can be reconstructed from arcminute-scale $B$-mode measurements, and the 
expected lensing contamination to degree-scale $B$ modes predicted and subtracted 
from the observed $B$-mode map. This procedure requires arcminute-resolution $B$-mode surveys 
(which in turn require large ($>$1m) primary 
apertures), but it complements debiasing because very different assumptions are made in 
the lensing removal process. 

For simplicity we consider a survey with uniform coverage at a given angular resolution.
The degree-scale and arcminute-scale measurements are provided by the same Stage-IV experiment: 
an array of platforms
each fed by $>10k$ background-limited detectors.  We use the \planck\  Sky Model (PSM) to predict the level of foreground 
contamination at various frequencies.  For each $\fsky$ (fraction of sky covered) considered, the cleanest patch is 
identified in terms of foreground level at 95 GHz, which is near the minimum in 
foreground contamination vs.~observing frequency \cite{dunkley09}.  In lieu of map-based component separation, 
we assume a level of residual foregrounds corresponding to $10\%$ or $5\%$ of the PSM. It is expected that this level of foreground 
removal will be achievable with good frequency coverage.  We follow the prescription in \cite{Smith09} to predict 
residual lensing contamination after ideal maximum-likelihood delensing \cite{Seljak:2003pn}.  We conservatively 
remove the first Fourier bin in each $\fsky$ considered to simulate loss of information caused by map 
boundaries.  

Results of these forecasts for a few representative Stage-IV configurations are shown in Figure \ref{fig:tensor}.
The goal of $\sigma(r)=10^{-3}$ is achieved or exceeded for a variety of configurations.
The constraint relies more significantly on delensing for smaller sky 
fractions and on debiasing for larger sky fractions.  Both methods are equally 
valid, and as a result, the formal $\sigma(r)$ evaluated at the optimal $\fsky$ 
is relatively insensitive to the resolution of the experiment: for an experiment with 
$1.1 \muk \sqrt{\mathrm{s}}$ sensitivity, $\sigma(r)$ 
improves modestly from $2.9\times 10^{-4}$ to $1.8\times 10^{-4}$ when the beam size decreases from $8'$ to $1'$ FWHM.  
However, it is important to point out that only delensing enables high S/N mapping of the inflationary 
$B$ modes if $r<0.02$, a level below which debiasing can only provide statistical measurements.

\subsection{Measuring the Spectrum of Primordial Density Fluctuations}
\label{powspecintro}
While the tensor-to-scalar ratio $r$ is in many ways the holy grail of inflationary CMB
science, it is not the only constraint on inflation models provided by the CMB. 
The dynamics of inflation also leave an imprint on the power spectrum of scalar 
density fluctuations. The simplest inflation models predict a nearly scale-invariant
power-law spectrum, and the most basic observational information about the primordial 
power spectrum is the power-law slope. 

Inflation is a nearly time-translation invariant state; however this invariance must be broken 
for inflation to eventually come to an end. In the inflationary paradigm, the wavelength of 
perturbations depends solely on the time that they were produced, thus a time-translation 
invariant universe would produce scale-invariant perturbations ($n_s = 1$, where the
scalar power spectrum is parameterized as $P(k) = A(k_0) (k/k_0)^{n_s-1}$).\footnote{Scale 
invariance here means that the contribution to the rms density fluctuation from a logarithmic 
interval in $k$, at the time when $k = aH$, is independent of $k$.  Here $a(t)$ is the scale 
factor and $H\equiv \dot{a}/a$ is the Hubble parameter.} The prediction that inflation should 
be nearly, but not fully, time-translation invariant gives rise to the prediction that $n_s$ should 
deviate slightly from unity. Recent CMB results have strongly confirmed this prediction, 
measuring $n_s = 0.96 \pm 0.007$---i.e., very close to but significantly departing from 
the scale-invariant value of $n_s=1$. In addition to confirming a prediction of inflation, 
these measurements, combined with upper limits on $r$, have begun to narrow the 
inflationary model space. A classic example with a potential  $V \propto \phi^2$ lies at the 
boundary of the $2\sigma$ region allowed by current CMB data.

With even more sensitive measurements of the temperature and $E$-mode polarization
of the CMB, we can go beyond this simple parametrization of the scalar power 
spectrum and investigate its detailed shape. The next-order measurement will be the
deviation from a power-law spectrum, parametrized as 
\begin{equation}
n_s(k) = n_s(k_0) + dn_s / d \ln k \ln\left(\frac{k}{k_0}\right).
\end{equation}
The value of $dn_s / d \ln k$ is often referred to as the ``running" of the scalar spectral
index.
The running parameter is predicted to be undetectable by most inflationary 
theories, and a detection of non-zero running could provide information about the 
inflationary potential or point to models other than inflation. The Stage-IV experiment
described in the previous section should be able to 
significantly improve the current constraints on running, particularly
through an exquisite measurement of the $E$-mode damping tail.
Similar measurements can be made using clustering of galaxies in
combination with CMB data. 
Current measurements from
\planck\  \cite{Ade:2013uln} indicate $dn_s / d \ln k = -0.0134 \pm 0.009$.
For \planck\  in combination with BOSS and
eBOSS, we forecast error on running of $0.006$ improving to $0.004$
for combination with DESI. Forecasts for how well the Lyman-$\alpha$
forest measurements will do are highly uncertain and depend on the
modeling advances but we are predicting error of $0.002$ for DESI
Lyman-$\alpha$ forest measurements.
These errors are larger than naturally predicted by
inflation, so new physics will have to be at work if we detect running
at these levels.  Experiments sensitive to running will also have the
power to search for features in the power spectrum beyond the first
few terms of a Taylor expansion. Such features are predicted in
certain inflation models with sharp features in the inflationary
potential or interactions during inflation.

\subsection{Beyond the Power Spectrum: Constraining Inflation through Higher-order Correlations}
\label{nongaussintro}
As discussed in Section \ref{sec:png}, 
in the inflationary paradigm, microscopic quantum fluctuations are the 
seeds of all the structure we see in the CMB and traced by collapsed objects like galaxies and 
clusters of galaxies. Due to the nature of these quantum fluctuations (specifically their
random phase), we expect the resulting distribution of density perturbations in the 
post-inflation universe to be very nearly Gaussian (as long as they are still in the
linear regime).  In particular, the standard class of single-field, slow-roll inflation models 
predict a level of non-Gaussianity of $\fnl^{\rm equilateral} < 1$ \cite{Maldacena:2002vr, Creminelli:2003iq} and $\fnl^{\rm local} = 0$ \cite{Pajer:2013ana}.
However, many models of inflation do predict a detectable level of 
non-Gaussianity. This means that a detection of primordial non-Gaussianity would 
rule out single-field slow-roll models, and that increasingly tight upper limits
will rule out many other models.

\subsubsection{Non-Gaussianity from the CMB}

The current best limits on primordial non-Gaussianity are obtained using data from the \planck\  satellite \cite{Ade:2013ydc}:
$\fnl^{\rm local} = 2.7 \pm 5.8$, $\fnl^{\rm equilateral} = -42 \pm 75$ and $\fnl^{\rm orthogonal} = -25 \pm 39$.  At the angular scales that
contribute most of the weight to the $\fnl$ constraints, \planck\  has measured the 
CMB temperature fluctuations as well as they can be measured (i.e., the constraints
on $\fnl$ is now limited by cosmic variance, not noise). Adding CMB polarization 
information will improve this constraint, but at most by $\sqrt{3}$.

\subsubsection{Non-Gaussianity from Large-Scale Structure Measurements}

Non-Gaussian features in the primordial fluctuations modulate the
subsequent evolution of structure in the universe and give rise
to effects that are present in the statistics of the density field at all redshifts.
The signature of non-Gaussianity is therefore expected to appear in the
abundance of massive clusters and in the large scale distribution of galaxies
and other tracers of dark-matter fluctuations.
Large spectroscopic programs such as BOSS
measure the LSS field in three dimensions over a
very large comoving volume, and thus offer a considerably
larger number of modes than the CMB.
The suite of large-area, spectroscopic BAO surveys
expected to occur between 2009 and 2022 (BOSS, eBOSS and DESI)
will therefore provide valuable constraints that will complement
the current \planck\  bispectrum limits and future CMB limits on the tensor to scalar ratio.

As opposed to the CMB, where non-Gaussianity is derived primarily from
considering higher-order correlators, in wide-field optical
spectroscopic surveys, non-Gaussianity can be constrained through the
two-point function. The easiest way to understand this is to consider
the tracer formation process as a non-linear transformation that
brings higher-order moments into the two-point function. In
particular, it has been shown that the typical tracers of cosmic
structure produce a quadratic divergence in power at large scales that
is proportional to $f_{\rm NL}^{\rm local}$ \cite{Dalal:2007cu}. This
has been used several times to place independent constraints on
non-Gaussianity (see e.g.,
\cite{Slosar:2008hx,Giannantonio:2011ya,Xia:2011hj,Ross:2012sx,Giannantonio:2013uqa}
and references therein). However, even the most competitive limits
from this technique ($-37 < f_{\rm NL}^{\rm local} < 25$ in
\cite{Giannantonio:2013uqa}) are dwarfed by the \planck\ results.
Predictions for BOSS, eBOSS and DESI indicate 68\% confidence limits
$\Delta f_{\rm NL}^{\rm local}=$24, 12 and 5, respectively
\cite{Font-Ribera:2013rwa}.  While there is some scope for improvement
with better techniques that reduce the impact of sample variance
(\cite{Seljak:2008xr}), the non-Gaussianity limits from galaxy
power-spectra remain useful mostly as a completely independent
cross-check of CMB limits.

Non-linear evolution of the dark-matter fluctuations
and the details of how sources trace the underlying field give rise to
additional, non-primordial, sources of non-Gaussianity in the evolved galaxy field.
This evolution must obey certain physical constraints (for example,
gravitational evolution and redshift space distortions can only
affect certain triangles of the 3-point correlation function).
Due to these constraints on non-linear density growth,
information from primordial non-Gaussianity is preserved even at smaller scales.
Higher-order
statistics such as the 3-point and 4-point correlation functions 
should provide superior constraints to those
derived from the power spectrum.  Constraining these correlation functions is especially important for equilateral and orthogonal type non-Gaussianity, 
which cannot be tested through the power spectrum \cite{Sefusatti:2007ih}.
While we do not have a reliable code to project  
non-Gaussianity constraints in this way, it is believed that these
should provide constraints at least as competitive as the best
projected CMB constraints,
provided that systematics in target selection and uniformity of 
spectroscopic observations over large areas can be brought under control.
We note however, that no constraint on non-Gaussianity from
higher order statistics has yet been demonstrated with current data.

\section{Conclusions}
The next generation of CMB and LSS experiments are poised to dramatically increase
our understanding of fundamental physics and the early universe by probing the
inflationary epoch. 
In particular, constraints on the amplitude of tensor modes will provide
unique insight into the physics of inflation only available through CMB observations.
A Stage-IV CMB experiment such as \cmbexp\ that surveys $> 1\%$ of the sky
to a depth of $\sim 1 \ \muk$-arcmin will deliver a constraint on the tensor-to-scalar ratio $r$
that either will result in a $5 \sigma$ measurement of the energy scale of inflation or
will rule out all large-field inflation models, even in the presence of foregrounds and the 
gravitational lensing $B$-mode signal. Such an experiment is technically feasible on
the timescale envisioned in this document.
The next generation of LSS measurements will complement the CMB effort 
by providing improved constraints on the cosmological
parameters associated with inflationary models.
Including all spectroscopic surveys through DESI, LSS measurements are expected
to improve current constraints on running of the spectral index by up to a factor of
four, improve constraints on curvature by a factor of ten, and provide non-Gaussianity constraints that are competitive
with the current CMB bounds.

%\vspace{5 mm}
\clearpage

\textbf{\em Note Added in Proofs:} 

After this paper was submitted, the BICEP2 collaboration reported a detection of $B$-mode polarization 
at degree angular scales \cite{Ade:2014xna}. When interpreted as an inflationary gravitational-wave
signal, the BICEP2 measurement implies a value of the tensor-to-scalar ratio of $r=0.2$. This result
immediately shifts the focus of currently operating and funded CMB experiments to confirming the level of the 
BICEP2 measurement and its cosmological origin, and it strengthens and clarifies the case for
a Stage IV CMB experiment. If the large value of $r$ is confirmed, the optimal \cmbexp\ survey strategies 
for the neutrino and inflation science goals become more closely matched: both science goals now 
drive the survey design to large sky areas to reduce sample variance. This level of tensor fluctuations
also makes the tensor spectral index $n_t$ a realistic target, making de-lensing an even more critical 
component of the \cmbexp\ design.

\bibliography{bibCF5}{}
 
%\input Magnetism/wgreport.tex 
%\input ChargedLeptons/wgreport.tex
%\input HeavyPhotons/wgreport.tex

%%%%%%%%%%%%%%%%%%%%%%%%%%%%%%%%%%%%%%%%%%%%%%%%%%
%%%%%%%%%%%%%%%%%%%%%%%%%%%%%%%%%%%%%%%%%%%%%%%%%%
%%%   Your subdirectory (here CF0) should include
%%%    the files:
%%%           wgreportCF0.tex
%%%           authorlistCF0.tex
%%%           bibCF0.bib
%%%         and all needed figures in pdf format
%%%%%%%%%%%%%%%%%%%%%%%%%%%%%%%%%%%%%%%%%%%%%%%%%%%%
%%%%%%%%%%%%%%%%%%%%%%%%%%%%%%%%%%%%%%%%%%%%%%%%%%%%

\end{document}